\documentclass[aps,prd,superscriptaddress,eqsecnum,amsfonts,showpacs,epsf]{revtex4}
\usepackage{epsfig}

\newcommand{\be}{\begin{equation}}
\newcommand{\ee}{\end{equation}}
\newcommand{\bea}{\begin{eqnarray}}
\newcommand{\eea}{\end{eqnarray}}

\newcommand{\nn}{\nonumber}

\begin{document}


\title{Analytical solution for the correlator with Gribov propagators}

\author{V. \v{S}auli}

\email{sauli@ujf.cas.cz}
\affiliation{Department of Theoretical Physics, Institute of Nuclear Physics Rez near Prague, CAS, Czech Republic  }

\begin{abstract}
Propagators approximated by a meromorphic functions with complex conjugated poles are widely used to
model infrared behavior of QCD Green's functions. 
In this paper,  analytical solutions for two point correlator made out of functions with complex conjugated poles or branch points have been obtained in the Minkowski space for the first time. As a special case the Gribov propagator has been considered as well.
The result is  different from the naive analytical continuation of the  correlator primarily  defined in the Euclidean space.
It  is free of  ultraviolet divergences, and instead of Lehmann  it rather  satisfies Gribov integral representation.   

\end{abstract}

\maketitle


%
\section{Introduction}

The explanation  of hadron properties in terms of QCD degrees of freedom represents hard non-perturbative task, especially  when 
the energy of a process does not comply with the asymptotic freedom and particle like description of hadron constituents.
Chiral symmetry breaking and confinement are the main phenomena beyond the applicability 
of perturbation theory. 

That confinement can be naturally encoded in  analytical properties of QCD Green's functions is an 
old-fashionable conjecture   \cite{GRIBOV1978,STINGL1986,ZWAN1989,MAHO1992,ZWAN1993,STINGL1996,DGSVV2008,BDGHSVZ2010,DG2011}.
Quark and gluon propagators with complex conjugated singularities are  
longstanding outcomes of many Bethe-Salpeter and Schwinger-Dyson equations (SDEs) studies \cite
{SC1990,BPT2003,DOKAKA2014,DKHK2014,HPGK2014,FKWa2014,ROBERTS}, noting here that a considered 
propagators, either calculated or phenomenologically used, usually exhibit not one but  several (infinite numbers with zero measure i.e. the cuts is possible as well) complex conjugated poles.
For instance the quark propagator considered as a series
\be   \label{propag}
S_q=\sum_i \frac{r_i}{\not p-m_i}+\frac{r_i}{\not p-m_i^{*}}  \, ,
\ee
where $m_i$ are complex numbers and  $r_i$ real residua, actually provides a good ingredient for calculations of 
pion observables. Note that in practice, most of (and all cited above) non-perturbative studies are based on the use of Euclidean metric from the beginning and the calculations performed in Minkowski space are due to the known obstacles  very rare
\cite{PRAROS2001,SAU1}.  Guiding by simple  assumption, that physics in  Minkowski space can be read from the analytical continuation of the solutions via  Euclidean theory, the lattice data has been checked against the form of Stjieltjes representation \cite{DUOLSI2014,SODB2014} as well as the solutions of SDEs system \cite{SAULI2012,SAULI2006,SAULI2007} have been discussed in the context of usual dispersion relation.

Assumed  structure of  propagators describing the confinement of quarks and gluons, 
i.e. the form represented  by (\ref{propag}),  implies the lost of perturbative analyticity.
 In this paper we  readdress some important issues of  analytical properties of Green's functions in QFT. For this purpose we consider 2-point correlation function of the following form
\be \label{masto}
\Pi(p)=i\int\frac{d^4l}{(2\pi)^4} \Gamma G(l) \Gamma G(l-p) \, .
\ee
Using for this purpose the  Gribov propagator:
\be \label{kokso}
G(p^2)=\frac{1}{p^2+\frac{\mu^4}{p^2}} \, \, ,
\ee
(here we ignore the spin structure) we will explicitly calculate the correlator (\ref{masto})  in Minkowski space and show that the correlator does not satisfy spectral representation but it reflects,  and in fact it reproduces again, the Gribov form in its own continuous  integral representation. Recall that  the lost of reflection positivity is expected for theory with confinement.
Here this is the specific form of analyticity, which ensures non existence of spectral (Lehmann) representation at all.

The choice of  the function (\ref{kokso}) is well  motivated due to  the fact that the Gribov propagator  appears as
 the solution of Gribov copy problem \cite{GRIBOV1978,ZWAN1989}. Actually, it was shown in \cite{CAP2015,PER2015,DEK2015} that if QCD is properly quantized then the gluon propagator in Landau gauge receives the Gribov form in non-perturbative manner. Notice trivially, it has the simplest nontrivial complex conjugated poles, which are located on the imaginary axis of $p^2$ complex plane.

The purpose of presented paper is twofold: the first point is to show that the evaluation of the loop with the propagators 
(\ref{kokso},\ref{propag}) in   Minkowski space is feasible and its analytical evaluation is certainly not too much demanding. 
As the second point, we mention possible physics related with. 

 Of course, elaborating a correct analytical relations between relations in  Minkowski and Euclidean spaces
is always useful and obviously possible here. 
However let us stressed, that we start with Minkowskian definition and thus the final result is not given by analytical continuation of the expression primarily defined within the use of a real Euclidean theory.
As a consequence, we will show that the correlator with Gribov propagators produces purely imaginary result, contradicting  not only the use naive Wick rotation, but the usual strategy. Furthermore, the obtained result is ultraviolet finite, which  is  another striking outcomes of presented calculation performance.

Regarding two point correlator matrix (\ref{masto}), it has
played longstanding and historically unpredentecable role in physics. Such two point function can stands for QED like  $V-V$ or $A-A$ current correlators considered originally  in  QCD Sum Rules \cite{SUM1,SUM2} or it can represent colored  gluon polarization function, quark selfenergy etc. depending on what we mean by the vertex $\Gamma$ and the propagator(s) $G$. 
While the  results do not support too much the old -fashionable conjecture of quark-hadron duality,  
the  integral representation derived thorough this paper can be actually used in practice for bound states and form factors calculations. Remind here its weak coupling perequisitor: the Perturbation Theory Integral Representation \cite{PTIR}, which  has found its own application in bound state calculations \cite{sauada2003,frsavi2012} in nonconfinig theory.

  We will consider not only Gribov propagator, but a wider class of the functions which have the complex conjugated singularities, including the poles as well as branch point in general. 
Before going ahead we simplify and concern  the analytical structure alone. 
However note, that using suitable "trace projectors" and after some trivial algebra, any  correlator matrix (\ref{masto}) can be cast into the sum of the product of matrices,tensors and  the following scalar form factors
\be \label{cor}
\Pi(p^2)=iI(p^2)=i\int\frac{d^4l}{(2\pi)^4} \Gamma(l,p) G(l)G(l-p)\, ,
\ee 
where all the functions in Rel. (\ref{cor}) are Lorentz scalars. Thorough this paper,  we neglect also the momentum dependence of the vertex function and simply take $\Gamma=1$, implying thus, that all the analytical behaviour  is  solely due to  the Minkowski space measure and the form of  functions $G$. Obviously, a more complete solution of the SDEs with nontrivial momentum dependence of the vertex would require analysis beyond the scope of presented study.
In what follow we present Minkowski space calculation for the  correlator (\ref{cor})
made out of Gribov propagators in the next Section (\ref{grib}). 

In order to see the effect of changing the pole position,  
we also consider the  correlator made out of the propagators with shifted poles.
 For this purpose we consider the following superconvergent toy model propagator function
\be 
G(l)=\frac{1}{(l^2-a)^2+b^2} \, ,
\ee
where $a$ represents  a real part of the pole location.
As the last case we consider also the convolution of "propagators" which have complex conjugated branch points. 
The later example with the function $G$ defined as 
\be \label{alf}
G(l)=\frac{1}{\sqrt{(l^2-a)^2+b^2}} \, ,
\ee
is considered in the Section IV.

\section{Correlator with Gribov propagators} \label{grib}.

The Gribov propagator (\ref{kokso}) 
represents a  simple rational function, whilst it has  a usual perturbative ultraviolet asymptotic,  
its infrared structure is drastically different from the free particle  propagator.
Instead of the real pole associated with free particle modes, it has  two simple  imaginary  poles associated with the confinement scale $b$. Its reality and the absence of the real axis  singularity  implies that  the function $I=-i\Pi$  should be real again. Also, the direct integration in the Minkowski space is well established  without a need  of sometimes unavoidable  analytical continuation to the auxiliary Euclidean space.
However, as  the  momentum integration is particularly easy within  the use of the Euclidean metric, the method of analytical continuation still remains a powerful technical  tool and we will use it carefully in our case as well. 
Before starting doing so, it is convenient to make a little algebra and we rewrite 
 the correlator $\Pi(p^2)$ in the following way:
\bea \label{gribci}
\Pi(p^2)&=&\frac{i}{4}\int\frac{d^4l}{(2\pi)^4}
\left[\frac{1}{(l^2+ib)(q^2+ib)}+\frac{1}{(l^2-ib)(q^2-ib)}\right.
\nn \\
&+&\left.\frac{2}{(l^2+ib)(q^2-ib)}\right] \, \, .
\eea 
where each line in bracket is hermitean and $q=l-p$.
Feynman parametrization represent a useful  tricks, which allows to evaluate perturbative loops in closed form.
  Using  this we can  write for the first line in (\ref{gribci})
\be \label{carka}
i\int\frac{d^4l}{(2\pi)^4}\int_0^1 dx \left[\frac{1}{[l^2+p^2x(1-x)+ib]^2}
 +\frac{1}{[l^2+p^2x(1-x)-ib]^2}\right]
\ee
where, we assume the shift of variable $l$ as well as the interchange of the integrations order is granted by the symmetry (and we will add omitted  factor $1/4$ in (\ref{gribci}) at the end).
Considering $l_o$ integration we will use the usual contour, which is  literally known as  a Wick rotation (WR) for the first term, while 
we will use the contour which is mirror symmetric to the Wick rotation (MWR) to integrate the second term (for Wick rotation see the standard textbook (\cite{textbook}) . Cauchy lemma then allows to write
\bea \label{koblizek}
\int\frac{d^4l_E}{(2\pi)^4}\int_0^1 dx\left[ \frac{-1}{[-l_E^2-p_E^2x(1-x)+ib]^2}
 +\frac{1}{[-l_E^2-p_E^2x(1-x)-ib]^2}\right] \,.
\eea
where $-p^2\rightarrow p_E^2=p_4^2+p_1^2+p_2^2+p_3^2$ in our metric convention.
 
Matching two terms in (\ref{koblizek}) together in the following manner
\bea 
&&\int\frac{d^4l_E}{(2\pi)^4}\int_0^1 dx \frac{4ib(-l_E^2-p_E^2x(1-x))}{[-l_E^2-p_E^2x(1-x)-ib]^2[-l_E^2-p_E^2x(1-x)+ib]^2}
\nn \\
&=&\int\frac{d^4l_E}{(2\pi)^4}\int_0^1 dx \int_0^1 dy\frac{\Gamma(4) 4ib y(1-y)(-l_E^2-p_E^2x(1-x))}{[l_E^2+p_E^2x(1-x)+ib(1-2y)]^4} \,.
\eea
The result is obviously finite and one can integrate over the momentum $l$ without  use of any regulator.
However it requires the integration over two auxiliary variables $x,y$  and there is a 
slightly easy way, which is to integrate each term individually. This step requires some usual (translation and other symmetries keeping) regularization, however the infinite peaces cancel each other. Independently on the procedure, it leads to the result:
\be
\frac{1}{(4\pi)^2}\int_0^1 dx
\ln\left(\frac{p^2_E x (1-x)+ib}{p^2_E x (1-x)-ib}\right) \, .
\ee

In order to get the analytical structure more explicit , it is advantageous to have all logs linearly dependent on integral variable.
An easy way is to exploit the substitution $u=x-1/2$ as a first. Then , realizing the integrand is the even function of $u$
one can  change the u-integral boundaries such that $\int_{-1/2}^{1/2}du\rightarrow 2\int_{0}^{1/2}du$. Then changing variable $u\rightarrow \omega$  such that $u=\sqrt{(1/4-b/\omega)}$ one  gets 
\bea
&&\frac{b}{(4\pi)^2}\int_{4b}^{\infty}\frac{d\omega }{\omega^2 \sqrt{\frac{1}{4}-\frac{b}{\omega}}}
 \ln\left(\frac{p^2_E +i\omega}{p^2_E -i\omega}\right)
\nn \\
&=&\frac{-4ib}{(4\pi)^2} \int_{4b}^{\infty}\frac{d\omega }{\omega^2 \sqrt{1-\frac{4b}{\omega}}}\tan^{-1}
\left(-\frac{\omega}{p^2}\right) \label{rammstein}
\eea
The expression (\ref{rammstein}) defines function of $p^2$ which has two symmetric cuts along imaginary axis  going from 
$i4b$ to $i\infty$ and from $-i4b $ to $-i\infty$.

Note that here is no real cut associated with the particle threshold and the usual of dispersion relation between absorptive and real part does not apply there. In other words: there is no Lehmann representation  for the correlator made out of Gribov propagators in Minkowski space. The correlator has no real part anywhere for the  real Minkowski space argument $p^2$. 
In formal analogy with perturbation theory which deals with the usual particle like propagators, an analogous  integral representation to spectral ones  can be written down, however here, they have Gribov form  again (i.e. denominator of such representation  has complex conjugated zeros). To write down such representation explicitly  one can use per-partes integration in $\omega$ variable
getting the following:
\be \label{Rela1}
\frac{i}{(4\pi)^2}\left[\pi \,  {\mbox sgn}{(p^2)}-2\int_{4b}^{\infty}d\omega\frac{ p^2 \,  \sqrt{1-\frac{4b}{\omega}}}{{p^2}^2+\omega^2}\right]\, ,
\ee
which shows us how the spectral representation for particles turns to the continuous sum of Gribov propagators 
of confined objects.  Anticipate here, the same arguments apply   for the remaining term in Eq. (\ref{gribci}) for which we are going to derive the appropriate result now. Matching together its  denominators by using Feynman
 variable $x$ we can write for the second line of Rel. (\ref{gribci})
\be
i\int \frac{d^4 l}{(2\pi)^4}\int_0^1\frac{2}{\left[l^2+p^2x(1-x)+ib(2x-1)\right]^2} \, .
\ee
To arrive to the known Euclidean integral we use $x-$parameter dependent contours in the complex $l_o$ plane. 
For $b(2x-1)$ positive (negative) we can use WR (MWR) and employ Cauchy  lemma (assuming the same for the external momentum).
This directly gives the following result:
\be \label{separat}
2\int \frac{d^4 l_E}{(2\pi)^4}\int_0^1dx\frac{\left[-\Theta(2x-1)+\Theta(1-2x)\right]}{\left[l_E^2+p_E^2x(1-x)-ib(2x-1)\right]^2} \, .
\ee
Integrating over the Euclidean momentum we arrive into the finite expression
\be
\frac{2}{4\pi^2}\left(-\int_{1/2}^1dx+\int_0^{1/2}dx\right)
\ln\left(-p_E^2 x(1-x)+ib(2x-1)\right) \, ,
\ee
noting that separate integration over  the first or over the second step function in  Eq. (\ref{separat})  would require  the introduction of  some regulator method  (like in the previous case, the regulator can be avoided by  summing these terms before the  integration, for which purpose one can use Feynman trick once again. We are not showing these trivial details).

In addition we offer several integral representations, which can be in principle useful in future. To arrive in, we  make the substitution $x=(1+z)/2$($x=(1-z)/2)$) in the first (the second) term. One immediately  gets
\bea \label{infected}
&&\frac{1}{(4\pi)^2}\int_0^1 dz
\ln\frac{-p_E^2 (1-z^2)/4-ib z}{-p_E^2 (1-z^2)/4+ib z}
\nn \\
&=&\frac{2i}{(4\pi)^2} \int_0^1 dz \tan^{-1} \frac{4bz}{p_E^2(1-z^2)} \, .
\eea

As aforementioned , it cannot be cast into the form of spectral representation.
The reason is obvious , the correlator has branch cut along the imaginary axis of $p^2$ instead of the real one.
Actually, using the substitution $z\rightarrow \omega$  one gets for (\ref{infected})
 the following representation
\be  \label{Rela2}
\frac{i}{(4\pi)^2}\left[-\pi \, {\mbox sgn}(p^2)+\int_0^{\infty}\frac{d\omega}{({p^2}^2+\omega^2)}\frac{p^2\omega}{(b+\sqrt{b^2+\omega^2/4)}}\right]
\ee
or alternatively
\be 
\frac{1}{(4\pi)^2}\left[-i\pi \, {\mbox sgn}(p^2)-\frac{1}{2}\int_{-\infty}^{\infty}\frac{d\omega}{(ip^2-\omega)}\frac{|\omega|}{(b+\sqrt{b^2+\omega^2/4})}\right]\, .
\ee

To conclude, the correlator $\Pi$  is given by (quarter of) the sum of two contributions (\ref{Rela1}) and (\ref{Rela2}) 
and satisfies continuous "Gribov" integral representation 
\bea
\Pi(p^2)&=&i\int_0^{\infty}\frac{d\omega \, \rho_G(\omega) \, p^2}{{p^2}^2+\omega^2}
\nn \\
\rho_G(\omega)(4\pi)^2&=&\frac{1}{4}\frac{\omega}{b+\sqrt{b^2+\omega^2/4}}-\frac{1}{2}\sqrt{1-\frac{4b}{\omega}}\Theta(\omega-4b)
\eea
Obviously, likewise the Lehmann representation reflects the analyticity of Feynman propagator $G^{-1}=p^2-m^2+i\epsilon$,
the correlator made out two Gribov propagators copiously reproduces the analytical structure of Green's functions involved.

\begin{figure}[t]
\begin{center}
\centerline{  \mbox{\psfig{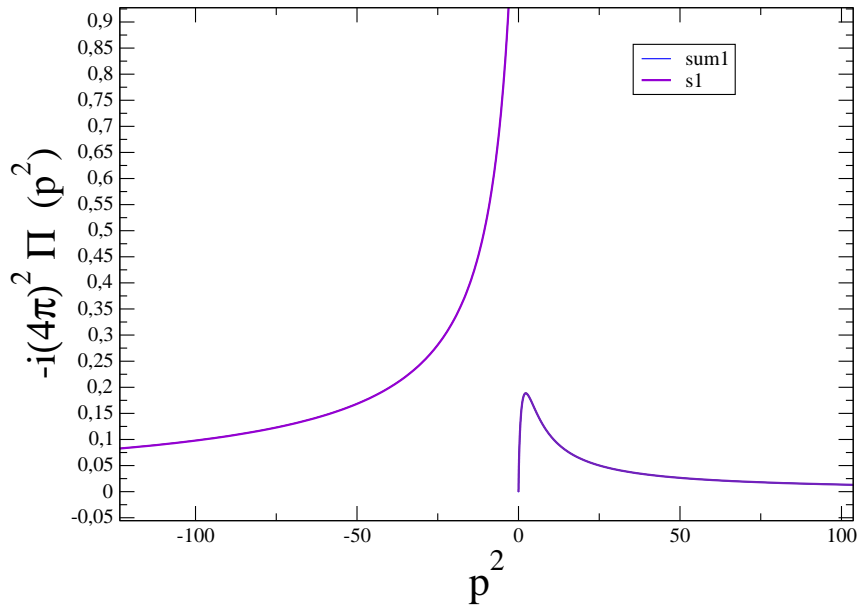}} }
\caption{Gribov correlator in units where $b=1$.\label{gribov0} }
\end{center}%
\end{figure}

The last undone  integration can be performed as well, a bit lengthy expression is  not presented here.
Numerical values for the function $\Pi$ is shown against the Minkowski variable $p^2$ in Fig. (\ref{gribov0}).
We plot separately  the result stemming from the first line for comparison. It is  shown that the terms which separately allows either WR or MWR  gives relatively smooth contribution in the infrared. While 
In contrast, there is  infrared discontinuity in  the origin of the complex $p^2$ plane, which stems from the second line, i.e. from the term which involves poles at both sides- upper and low semi-half- of complex plane.  
Recall, this discontinuity is lowered  when the  pole of Gribov propagator  is shifted away from the imaginary axes.
This analytical behaviour  is worthwhile to study and we do this in the next Section.

At last but not at least, we  comment the ultraviolet finiteness of the result. The result we obtained is finite here,
 however we should stress that ultraviolet finiteness  is not a general property of the correlator made out of propagators with complex conjugated poles. Usual "log" divergence arises for the convolution of two different Gribov  propagators. This divergence is proportional to the difference of the pole positions $\simeq b_1-b_2$ and if necessary it  can removed by some sort of symmetry keeping regularization. Recall here,  the Gribov form of QCD propagators is assumed to be a good approximation in the infrared, while it is likely less useful for the study of ultraviolet properties. However, as the selfconsistent and serious Minkowski space study of QCD with Gribov propagators is lacking, an outcomes can be challenging.     

\section{Correlators with generalized Gribov propagators}
\subsection{Shifting the branch point}

In the previous section we have derived Gribov integral representation, which arises when two Gribov propagators convolute in 3+1 momentum space. Obvious question arises: what would happen to the correlator  $\Pi$ when one consider more general structure of the propagator, e.g. shifted pole, branch points, etc. As an example we will  
 consider the  correlator $\Pi(p^2)$   
with propagators of the following form:
\be 
G(l)=\frac{1}{(l^2-a)^2+b^2} \, .
\ee

Again, without a detailed calculation one can see  obvious properties of the correlator. Like in previous case, the function  $\Pi(p^2)$  should be a purely imaginary function  for all $p^2$. Further obvious is its finiteness,  actually it should be finite  
since its Euclidean counterpartner is.

To arrive into the analytical expression here, we change a bit the calculation procedure and start with the Feynman parametrization from the beginning.  For this purpose let's  write 
\be
\frac{1}{(p^2-a)^2+b^2}=\frac{1}{p^2-a+ib}\frac{1}{p^2-a-ib}=\int_0^1 d x 
\frac{1}{[p^2-a+ib-2ibx]^2}     \, .
\ee
The  product of two  such propagators in (\ref{cor}) can be written as
\bea \label{sarka}
&&\int_0^1 dx_1 dx_2 \frac{1}{[l^2-a+ib-2 i b x_1]^2} \frac{1}{[(l-p)^2-a+ib-2 i b x_2]^2}
\nn \\
&=&\int_0^1 dy dx_1 dx_2 \frac{y(1-y) \Gamma(4)}{[l^2 y+(l-p)^2(1-y)-a+ i b - 2 i b(x_1 y+x_2 (1-y))]^4}
\eea
Lorentz invariance of the measure as well as  the  kernel here, both imply that the finite shift of the integral variable 
leaves the result invariant. We assume, this symmetry dictated property is a must of the theory, with 
possible anomalies exceptions which  we are not going to discuss here.    
Shifting  momentum $l$ one gets for the correlator:
\bea \label{Omega}
\Pi(p^2)&=&i\int \frac{d^4l}{(2\pi)^4}\int_0^1 dy dx_1 dx_2 \frac{y(1-y) \Gamma(4)}{[l^2+p^2(1-y)y-\Omega]^4} \, ,
\nn \\
\Omega&=&a- i b + 2 i b(x_1 y+x_2 (1-y))\, .
\eea
The position of the pole now depends on the parameters and we interchange  ordering of  integrations and integrate over the four-momentum 
as a first. Like in the previous case we use MWR for $\Im\Omega<0$ , while for positive $\Omega$, when the pole is located in the lower half plane of complex $l^2$, we will use the usual  WR. In both cases the inner part of  intended curve is thus free of singularities and the use of Cauchy lemma  switches  to the Euclidean metric (performing similar for the external momentum). Doing this explicitly, one gets the following prescription for the integral
\be \label{spaces}
i\int\frac{d^4 l}{(2\pi)^4} f(l,p) 
\rightarrow - \int\frac{d^4 l_E}{(2\pi)^4} \Theta(\Im \Omega) f(l_E,p_E)+\int\frac{d^4 l_E}{(2\pi)^4} \Theta(-\Im \Omega)f(l_E,p_E)\, ,
\ee
 wherein the subscript $E$ implies the arguments of $f_E$ uses an Euclidean metric $l_E^2=l_1^2+l_2^2+l_3^2+l_4^2$.  
In accordance with causality, we  assume $a>0$ in order to avoid poles at the spacelike region of momenta. 
Integrating over the momentum $l$ one gets 
\be
\Pi(p^2)=\int_0^1 dy dx_1 dx_2 \frac{y(1-y)\left[-\Theta(\Im \Omega)+\Theta(-\Im \Omega)\right]}{(4\pi)^2[p^2 y(1-y)-\Omega]^2} \, ,
\ee
 where $p$ is Minkowski  momentum again, and the function $\Pi(p^2>0)$  is an analytical continuation of $\Pi(p^2<0)$.

Performing the substitution$x_1\rightarrow z$ such that $z=1-2(x_1 y+x_2(1-y))$, one gets
\be
\Pi(p^2)=\int_0^1 dy dx_2 \int_{1-2(y+x_2(1-y))}^{1-2x_2(1-y)} dz 
\frac{(1-y)\left[-\Theta(z)+\Theta(-z)\right]}{(4\pi)^2[p^2 y(1-y)-a+ibz]^2} \, .
\ee
 
For the purpose of  the integration over the variable $z$ let's distinguish three cases. The first case we define, is such that upper and down $z-$integral boundaries are negative. This allows to consider only  the mirror Wick rotation. Integrating over the variable $z$ leads to the following formula:
\be \label{I1}
\frac{-1}{2ib}\int_0^1 dy dx_2 \frac{(1-y)}{(4\pi)^2} \Sigma_{i=1,2}  \frac{(-1)^{1+i}\left[\Theta(2x_2(1-y)-1)\right]}{p^2 y(1-y)-a+ib z_i} \, ,
\ee
 where $z_1=1-2x_2(1-y)$ and $z_2=1-2(y+x_2(1-y))$.

The second case is defined by inequalities $z_1>0$ and $z_2<1$. Both step functions are relevant  and they  just split the $z$ integration domain to two integrals with boundaries $z_2,0$  and $0,z_1$ respectively.   Integrating over the variable $z$ is straightforward and the result for the second case  reads
\be \label{I2}
\frac{-1}{2ib}\int_0^1 dy \int_0^1 dx_2 \frac{(1-y)\Theta(z_1)\Theta(-z_2)}{(4\pi)^2} 
\left[\frac{2}{p^2 y(1-y)-a}
-\Sigma_{i=1,2} \frac{1}{p^2 y(1-y)-a+ib z_i}\right] \, ,
\ee
with  two variables $z_i$ defined  previously.

The third and the ultimate case corresponds with the condition  $z_1>z_2>0$, for which one gets
\be  \label{I3}
\frac{-1}{2ib}\int_0^1 dy \int_0^1 dx_2 \frac{(1-y)\Theta(z_2)}{(4\pi)^2} 
\Sigma_{i=1,2} (-1)^i\frac{1}{p^2 y(1-y)-a+ib z_i} \, .
\ee
The scalar polarization correlator $\Pi$ is then equal to the sum of expressions (\ref{I1}), (\ref{I2}) and (\ref{I3}).

Further analytical integration is possible as well, for instance the integration  over the variable $x_2$ gives for (\ref{I2})
\bea \label{ucho}
&&\frac{1}{(2ib)^2}\frac{1}{(4\pi)^2}\left[\int_0^{1/2}d y \ln\frac{R-2 i by}{R+2 i by}
+\int_{1/2}^{1}\ln\frac{R+ i b(2y-1)}{R+ i b (1-2y)}
+\int_{1/2}^{1} \ln\frac{R+ i b}{R- i b}\right]
\nn \\
&+&\frac{i}{2b}\frac{1}{(4\pi)^2}\left[\int_0^{1/2}d y\frac{-y}{R}
+\int_{1/2}^1 d y \frac{1}{R}\right] \, ,
\eea
where we have defined $R=p^2 y(1-y)-a$ for purpose of brevity.

Using the identity
\be
\ln\frac{R-2 i by}{R+2 i by}=2i \tan^{-1}\frac{2 b y}{p^2y(1-y)-a}
\ee
for the first  and similarly for  other terms in (\ref{ucho}), one immediately  sees that the result for 
(\ref{I2}) is purely imaginary.

Further, summing (\ref{I1}) and (\ref{I3}) together and integrating over $x_2$ gives after some trivial algebra the following formula:
\be \label{212}
\frac{1}{(-2i b)^2}\frac{1}{(4\pi^2)}\int_0^{1/2} dy
\left[-\ln(R+2ib)+\ln(R+ib)-\ln(R+ib(1-2y))+\ln(R)+c.c.\right] \, ,
\ee 
where c.c. stands for complex conjugated term. Recall, as was discussed in the beginning,  the total result must be completely imaginary. Thus since Rel. (\ref{I2}) already is, while  the  Rel. (\ref{212}) is  purely real, the later must be exactly zero for all $p$ and 
the total contribution is  given solely by the expression (\ref{ucho}).

All integrals in (\ref{ucho}) can be further integrated analytically, providing the following final result:
\bea \label{final}
\Pi&=&\frac{i}{2b(4\pi)^2}\left\{\frac{1}{2s} \ln\frac{(a+s/4)^2+b^2}{a^2+b^2}-\frac{1}{4s} \ln\frac{(a-s/4)^2}{a^2}
+\frac{1}{2b} \tan^{-1}\frac{b}{a}-\frac{1}{s\sqrt{\frac{4a-s}{s}}}\tan^{-1}\sqrt{\frac{s}{4a-s}}\right\}
\nn \\
&-&\frac{1}{4b^2(4\pi)^2}\left\{\frac{\sqrt{D}}{s} \tan^{-1}\frac{-2ib+s}{\sqrt{D}}
-\frac{\sqrt{D}}{s}\tan^{-1}\frac{-2ib}{\sqrt{D}}-c.c.\right.
\nn \\
&+&\frac{\sqrt{D_1}}{s}\tan^{-1}\frac{2ib}{\sqrt{D_1}}-\frac{\sqrt{D_1}}{s} \tan^{-1}\frac{2ib-s}{\sqrt{D_1}}-c.c.
\nn \\
&-&\left.\sqrt{\frac{D_2}{s}}\tan^{-1}\sqrt{\frac{s}{D_2}}-c.c.\right\} \, .
\eea
where we have used the following abbreviations
\bea
D&=&s(4a-s)+4b^2 \, \, ;\, \,
D_1=s(4a-s)+4b^2+4 i b s \, \, ;
\nn \\
D_2&=&4a+4 i b-s \, ; s=p^2 \, \, .
\eea
The inverse tangent function  of complex argument is defined through the multi-valuable complex logarithm. As one can see, apart of complex singularities, there is a real branch point  presented as well. This branch point has coincided with the origin of the complex plane in the previous case, where $a=0$ case was considered only.

\subsection{$\Pi$ composed from propagators with  branch points}

Quark and gluon propagators can obtained via solution of SDEs
and the result can be in principle approximated  as a series (\ref{propag}).  
It is well known that the interaction is reflected in the analytical properties of amplitudes.
For instance, in a relatively simple theory like QED, the effect of dressing electron propagator by photon
self-exchange entails that a simple pole structure of the electron propagator turns to the branch point singularity. 
In QCD one can expect  the similar, if not stronger  effect and it is plausible that the other considerable expansions are much faster convergent , especially when their first terms catch the main properties of the exact solution. 
 As such a further reasonable candidate, we consider propagator with square root non-analyticity. Guiding by simplicity as well as it ultraviolet asymptotic of a free propagator, let us calculate the correlator with the propagator (\ref{alf}).
In this  special case, the argument of  the  correlator reads
\be  \label{sqrt}
\frac{1}{\sqrt{((l-p)^2-a)^2+b^2}}\frac{1}{\sqrt{(l^2-a)^2+b^2}} \, . 
\ee
As the first step we start with the root decomposition of  square-root arguments, then  using Feynman trick gives us
\be \label{rhsof}
\frac{1}{(p^2-a+ib)^{1/2}}\frac{1}{(p^2-a-ib)^{1/2}}=\int_0^1 d x 
\frac{x^{1/2}(1-x)^{1/2}\frac{\Gamma(1)}{\Gamma^2(1/2)}}{p^2-a+ib-2ibx}     \, .
\ee
Depending on the value of variable $x$, the denominator in rhs. of Eq. (\ref{rhsof}) has positive imaginary part for a small $x$ and positive $b$. It turns to be negative  for a larger value of $x$. Using the variable $x_1$ and $x_2 $ for each propagator in 
(\ref{sqrt}) and further using the variable $y$ to match propagators together one gets for Rel. (\ref{sqrt}) the following expression
\be
\int_0^1 \frac{dx_1\, dx_2\, dy \, \sqrt{x_1(1-x_1)x_2(1-x_2)}\frac{\Gamma(2)}{\Gamma^4(1/2)}}
{l^2y+(l-p)^2(1-y)-\Omega} \, ,
\ee 
where $\Omega$ is defined by (\ref{Omega}) in the previous Section. After the standard shift one gets for the correlator
\bea
\Pi(p^2)&=&i\int \frac{d^{3+1}l}{(2\pi)^{4}} \int_0^1
\frac{dx_1\, dx_2\, dy \, \sqrt{x_1(1-x_1)x_2(1-x_2)}\frac{\Gamma(2)}{\Gamma^4(1/2)}}{\left[l^2-\Delta\right]^2}
\nn \\
\Delta&=&-p^2(1-y)y+\Omega \, .
\eea

Now we will integrate over the momenta as in the previous case.  It means that for case when $\Im \Delta<0$ WR is used, while when the denominator has singularities at the first and the third quadrant of complex $l_o$  MWR is used. 
The conditions $\Im \Delta<0$ and $\Im \Delta>0$ limits the Feynman parameters integration domain.  The integration over the momentum, if taken separately, includes UV divergence.  Obviously, resulting infinite constants  cancel against each other at the end. Using dimensional regularization for this purpose one can write
\bea \label{logs}
\Pi(p^2)&=&\int \frac{1}{(4\pi)^2} \int_0^1
dx_1\, dx_2\, dy \, \sqrt{x_1(1-x_1)x_2(1-x_2)}\left(\frac{2}{\epsilon} -\ln(\Delta)-\gamma+O(\epsilon)\right)
\nn \\
&&\left[\Theta(-\Im \Omega)-\Theta (\Im \Omega)\right] \, .
\eea

\begin{figure}[t]
\begin{center}
\centerline{  \mbox{\psfig{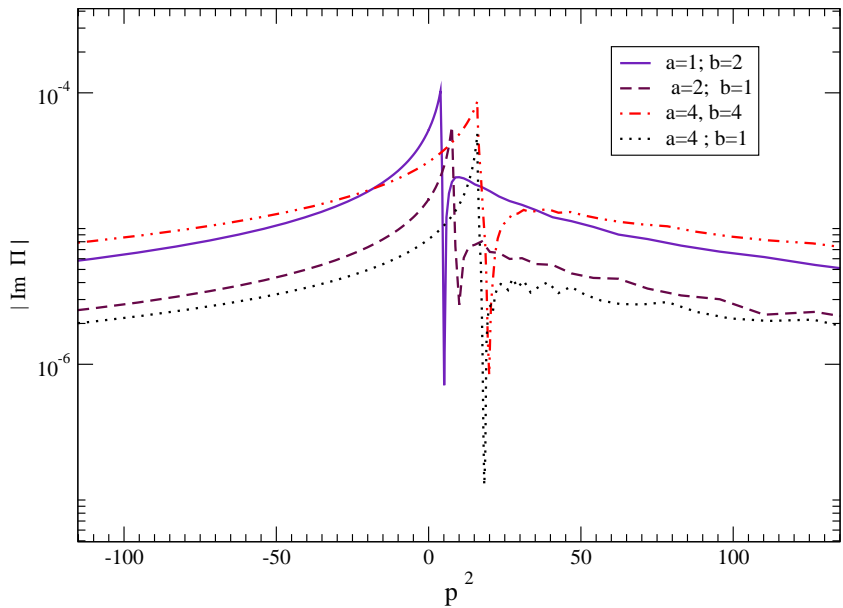}} }
\caption{Scalar crrelator made out  of the two propagators with branch point singularity. Momentum is scaled by the values of $a,b$, which are shown explicitly.\label{gribov1}}
\end{center}%
\end{figure}

Note that  only the $\Im$ parts of the expression in the large bracket in  Eq. (\ref{logs}) can survive at the end,  the result will not depend on the  renormalization scale at all and as mentioned it is finite at all. The remaining 3d integral over the Feynman variables can be performed numerically. If one wishes, the vanishment of the real part can regarded as test of numerical precision. Actually, numerically we get $\Re\Pi / \Im {\Pi} < 10^{-15}$ (here we did not find the analytical expression over the Feynman parameters due to the presence of square-root function).  The resulting correlator is plotted in Fig. (\ref{gribov1}). Obviously, one can see the evidence for a real branch point at the momentum  $p^2=4a^2$. 

\section{Conclusion}

Correlators defined as a convolution of  two  Green's functions with various analytical structure, which admits confinement,  have been studied in the momentum  Minkowski space. We have restricted to the  choice of real  propagators with complex conjugated singularities. The Gribov propagator, which plays important role in $SU(3)$ Yang-Mills theory 
was considered and studied in great details. It was shown that Feynman parametrization  allows  the analytical integration over the momentum in all studied cases, providing unique and ultraviolet finite results in the  Minkowski space.  
Contrary to  calculations performed in the Euclidean space \cite{DGSVV2008,BDGHSVZ2010,DG2011}, the Minkowski space correlator with Gribov propagators remains finite and does not require renormalization.  Neither of  correlators   satisfies Khallen-Lehmann representation and in  the case of Gribov propagator the integral representation  copiously reproduces the Gribov form in its continuous version. In other cases, the analytical structure is more complicated and there is no easy way to classify all branch points and related cuts. The correlator made out of generalized Gribov propagators exhibits   the  real quasithreshold as well.  The later  strikes itself as a sharp cusp in the graph of $\Pi(p^2)$ and is located at the usual point $p^2_T=4a^2$ ($p^2_T=0$ for Gribov).

The ramifications of the Minkowski space calculations can only be explored in  conjunction with quantum Chromodynamics in strong coupling region. In this respect many steps remain to be finished and even the Gribov form is not an ultimate approximation of QCD propagators.    
Meanwhile, there are  the first attempts \cite{SAU1,SAU2} to solve Schwinger-Dyson  and Bethe-Salpeter equations directly in the momentum Minkowski space, the presented study show the pertinent existence of discontinuities (cuts) when passing the real axis of momenta in Minkowski space. It obviously can make  the direct Minkowskian momentum space integration of SDEs difficult, cumbersome,  if not  even impossible in some cases. In this respect, the integral representation derived  here     can be very  useful when looking for hadronic observables in the framework of SDEs.


%
\end{document}